\documentclass[preprint,showpacs]{revtex4}
\usepackage{graphicx}
\begin{document}
\title{Scaling theory in a model of corrosion and passivation}
\author{F. D. A. Aar\~ao Reis$^1$, Janusz Stafiej$^2$ and J.-P. Badiali$^3$}
\affiliation{
$^1$Instituto de F\'\i sica, Universidade Federal Fluminense,\\
Avenida Litor\^anea s/n, 24210-340 Niter\'oi RJ, Brazil\\
$^2$Institute of Physical Chemistry, Polish Academy of Sciences,\\
Kasprzaka 44/52, 01-224, Warsaw, Poland\\
$^3$Universit\'e Pierre et Marie Curie, l'ENSCP, UMR 7575 du CNRS,\\
4. Place Jussieu, 75005 Paris, France}
\date{\today}

\begin{abstract}

We study a model for corrosion and passivation of a metallic surface after
small damage of its protective layer using scaling arguments and simulation.
We focus on the transition between an initial
regime of slow corrosion rate (pit nucleation) to a regime of rapid
corrosion (propagation of the pit), which takes place at the so-called
incubation time.
The model is defined in a lattice in which the states of the sites represent
the possible states of the metal (bulk, reactive and passive) and the
solution (neutral, acidic or basic). Simple probabilistic rules
describe passivation of the metal surface, dissolution of the passive layer,
which is enhanced in acidic media, and spatially separated electrochemical
reactions, which may create pH inhomogeneities in the solution. 
On the basis of a suitable matching of characteristic
times of creation and annihilation of pH inhomogeneities in the solution, our
scaling theory estimates the average radius of the dissolved region at the
incubation time as a function of the model parameters. Among the main
consequences, that radius decreases with the rate of spatially separated
reactions and the rate of dissolution in acidic media, and it increases with the
diffusion coefficient of $H^+$ and ${OH}^-$ ions in solution. The
average incubation time can be written as the sum of a series of
characteristic times for the slow dissolution in neutral media, until
significant pH inhomogeneities are observed in the dissolved cavity. Despite
having a more complex dependence on the model parameters, it is shown that the
average incubation time linearly increases with the rate of dissolution in
neutral media, under the reasonable assumption that this is the slowest rate of
the process. Our theoretical predictions are expected to apply in realistic
ranges of values of the model parameters. They are confirmed by numerical
simulation in two-dimensional lattices, and the expected extension of the theory
to three dimensions is discussed.
\end{abstract}

\maketitle

\section{Introduction}

The corrosion of a metal after the damage of its protective layer is a problem
of wide technological interest. The evolution of the
corrosion front is the result of a competition between localized dissolution
and passivation processes. The latter consists of the formation of a
passivation layer which reduces the corrosion rate and prevents a fast
propagation of the damage. However, the breakdown of the passivation layer
leads to the so-called pitting corrosion \cite{frankel}, with an increase in
the dissolution rate. It is generally accepted that the propagation of the pit
is preceded by a nucleation regime, and the transition between these regimes
takes place at the so-called pit initiation time or incubation time. This
characteristic time was experimentally observed for a long time
\cite{hoar,shibata}. In stainless steel, the process is often connected with
the presence of inclusions ($MnS$) from which the pits begin \cite{alkire}. In
weakly passivated materials, it often begins at surface defects and surface
inhomogeneities \cite{frankel}, with only a small fraction giving rise to
indefinitely developing pits. Some recent works determined relations between
the incubation times and physicochemical conditions in different processes of
technological interest (see. e. g. Refs.
\protect\cite{fukumoto,hassan,rehim,amin,zaky}).

There is also extensive literature covering the theoretical aspects of
pitting corrosion. In some cases, specific applications are considered, such as
stainless steel \cite{nagatani1,laycock1, laycock2}, while some works focus on
universal features of that process \cite{meakin1}. Among the models we may
also distinguish the ones based on stochastic approaches
\cite{nagatani1, meakin1, nagatani2, nagatani3, meakin2} from those based on
analytical
formulations of the corrosion problem \cite{laycock1, laycock2, digby,organ}.
These papers are devoted to the study of pit propagation, with the focus on the
pit shapes \cite{laycock1,meakin2,nagatani3} and their evolution or the
investigation of the interaction between pits \cite{organ}.

On the other hand, the transition between the nucleation regime and pit
propagation has attracted less attention. It motivated the recent study of a
stochastic model with mechanisms that may be involved in this transition
\cite{vautrin1,stafiej,passivity}.
These mechanisms are passivation/depassivation phenomena, generation of local pH
inhomogeneities by spatially separated cathodic and anodic reactions and
smoothing out of these inhomogeneities by diffusion. Simulation of the
corrosion process initiated by small damage to a protective surface showed
the existence of an incubation time separating a nucleation regime of slow
corrosion from a regime with a much higher growth rate of the corrosion front.
Qualitative features of the growing cavities in two-dimensional
simulations have already been addressed \cite{vautrin1,stafiej}. However,
a thorough analysis of the quantitative effects of different model parameters
is still lacking. 

In this paper, we extend the study of this corrosion model by combining scaling
ideas and simulation results. Since the number of parameters of the original
model is large, our study focuses on their values in the most realistic ranges
for possible applications.
Thus, while keeping the model amenable for a combination of theoretical and
numerical work, we attempt to preserve the perspective of applications.
The damage is represented by two lattice sites initially exposed to the
environment,
which mimics local damage of a relatively cheap (not stainless steel) painted
material in its natural environment. This is certainly a situation of
practical interest.

The starting point of our scaling analysis is to estimate
the characteristic times of the physicochemical mechanisms involved in the
corrosion process. By a suitable matching of those characteristic
times, we estimate the incubation radius, which is defined as the effective
radius of the dissolved region at the incubation time. Such a
procedure resembles those recently used
to estimate crossover exponents in statistical growth models 
\cite{rdcor,lam}. The incubation radius is related to model
parameters such as probabilities of spatially separated reactions (cathodic and
anodic), probabilities of acidity-enhanced depassivation and the diffusion
coefficients. Subsequently, the dependence of the average incubation
time on the model parameters is also analyzed.
The theoretical predictions are supported by simulation data for two dimensional
square lattices. This dimensionality is forced by computational limitations.
However, the physicochemical and geometrical arguments of the theoretical
analysis are independent of the underlying lattice structure. Thus we expect
that this independence extends to the results in a certain range of the model
parameters. This is important to justify the reliability of the model for a
mesoscopic description of corrosion phenomena.

The paper is organized as follows. In Sec. II we review the
statistical model and give a summary of the results of previous papers. In Sec.
III we present the scaling theory for the model, which predicts the average
radius of the dissolved cavity at the incubation time. In Sec. IV we compare
the theoretical predictions with simulation data. In Sec. V we discuss the
scaling properties of the incubation time. In Sec. VI we discuss the relations
of this model with other reaction-diffusion models and other models for pitting
corrosion. In Sec. VII we summarize our results and present our conclusions.

\section{The corrosion model}
\label {model}

\subsection{The electrochemical basis of the model}
\label{electrochemicalbasis}

In an acidic or neutral medium, the anodic dissolution of commonly used metallic
material such as steel, iron and aluminum, can be described by the reaction
\begin{equation}
Me + H_2O \to {MeOH}_{aq} + e^- + H^+ ,
\label{dissolacidic}
\end{equation}
where $Me$ represents the bulk metal material and ${MeOH}_{aq}$ represents
products which may be composed of hydroxides, oxides and water.
For example, the mechanisms of $Ni$ dissolution in acid phosphate solutions was
recently discussed in Ref. \protect\cite{munoz}. However, the precise
chemical nature of these products is not important for the model presented here,
but only the assumption that they are detached from the corroding material
and belong to its environment.

On the other hand, in a basic environment the surface is expected to
repassivate. In other words, the chemical species produced there are adherent to
the surface and compact enough to prevent further dissolution. We represent it
by
the reaction
\begin{equation}
Me + {OH}^- \to {MeOH}_{ads} + e^- ,
\label{dissolbasic}
\end{equation}
where ${MeOH}_{ads}$ refers to the adherent species forming the passive layer at
the metal surface.
In both cases the pH of the solution at the locus of the reaction
decreases. 

In acidic deoxygenated media the associated cathodic reaction is
\begin{equation}
H^+ + e^- \to \frac{1}{2} H_ 2 ,
\label{cathodicacidic}
\end{equation}
while in a basic environment we have
\begin{equation}
H_2O + e^- \to \frac{1}{2} H_ 2 + {OH}^- .
\label{cathodicbasic}
\end{equation}
As expected, these reactions increase the local pH.

If anodic and cathodic reactions occur next to each other there
is a mutual compensation of their effect on pH and eventual neutralization.
Thus the
pairs of reactions (\ref{dissolacidic})-(\ref{cathodicacidic}) (in neutral or
acidic medium) or (\ref{dissolbasic})-(\ref{cathodicbasic}) (in basic medium)
do not alter pH of the solution at a significant lengthscale.
Consequently, they can be simply combined as:
(\ref{dissolacidic}-\ref{cathodicacidic})
\begin{equation}
Me + H_2O \to {MeOH}_{aq} + \frac{1}{2}H_2 
\label{reacacidic}
\end{equation}
if the surrounding solution is acidic, or as
(\ref{dissolbasic}-\ref{cathodicbasic})
\begin{equation}
Me + H_2O \to {MeOH}_{ads} + \frac{1}{2}H_2 
\label{reacbasic}
\end{equation}
if the surrounding solution is basic.
Reaction (\ref{reacbasic}), in which the oxidation of the metal is followed by
the deposition of a passive species, is much more frequent than the
reaction (\ref{reacacidic}), in which the product of the corrosion immediately
leaves the surface. In a neutral medium, the former also tends to be dominant.

However, the spreading of the electric signal in the metal is considered as 
instantaneous when compared to the other processes. Therefore
oxidation (\ref{dissolacidic} or \ref{dissolbasic}) and reduction
(\ref{cathodicacidic}
or \ref{cathodicbasic}) may occur at distant points simultaneously. These
spatially separated
electrochemical (SSE) reactions change the local pH of the solution where they
occur
 in favor of occurrence of the reactions of the same type at the location. On
the other
hand, pH inhomogeneities (local excess of $H^+$ over ${OH}^-$) may be
suppressed by diffusion, which tends to reimpose uniformity and brings about
neutralization.

Finally, it is also important to recall that the dissolution of the adsorbed
species (${MeOH}_{ads}$ in the above reactions) is very slow in a neutral
medium, and even more difficult in a basic one. For instance, simulation of
corrosion on steel by $CO_2$  at a high pH \cite{nesic} shows low diffusion
rates,
with the formation of very protective films.
However, the dissolution is significantly
enhanced in acidic medium, following the reaction
\begin{equation}
{MeOH}_{ads} + H^+ \to {Me}^+ + H_2O .
\label{dissolaggressive}
\end{equation}
As the anions ${Me}^+$ leave the surface, the metal is again exposed to the
corrosion agents.

\subsection{The statistical model}
\label{statisticalmodel}

Here we rephrase the model of Ref. \protect\cite{vautrin1}, which
amounts to describing the above processes at a mesoscopic level. Metal and
solution are represented in
a lattice, whose sites may assume six different states. These states represent
the presence of significant amounts of the most important chemical species,
while the chemical reactions are represented by stochastic rules for the changes
of those states after each time step.

The  sites representing the bulk metal (unexposed to the corroding solution) are  
labeled M and called metal or M sites. The so-called R (reactive) sites
represent the
"bare" metal exposed to the corrosive environment. We consider the passivation
layer at these points permeable enough to allow for anodic dissolution of the
metal. In contrast, the passivated
regions, labeled P sites, represent sites covered with a passive
layer which is compact enough to prevent their anodic dissolution. The other
lattice sites, labeled E, A and B, denote the neutral
environment, the acidic and basic regions, respectively (compared to previous
work on this model \cite{vautrin1,stafiej,passivity}, label C is here replaced
by B in
order to make a clearer
association with the basic character of the solution and emphasize the pH
role).

The evolution of the corrosion process amounts to 
transformations of the interfacial sites (R and P) into solution sites (E, A or
B), followed by the conversion into R sites of those M sites that are put in
contact with the solution. This accounts for the displacement of
the interface. In the following, we will use the label S to denote a surface
site which can be either R or P.

The possible changes of site labels certainly depend on the pH of the
surrounding solution. Thus, for a given S site, our pH related scale will be
represented by the algebraic excess of A
sites over B sites among the nearest neighbors. We denote it by $N_{exc}$, so
that the pH decreases as $N_{exc}$ increases.

During the simulation of the process, SSE reactions, local vicinity reactions
and depassivation reactions are performed in this order. For each type of
reaction, the lists of R and/or P sites are looked up in random order, and the
decision of each one to undergo that reaction is taken with a prescribed
probability. These reactions are
followed by a certain number of random steps of A and B particles in the
solution and possible annihilation of their pairs. This series of events
takes place in one time unit.

The first set of transformations of the lattice sites account for the SSE
reactions. In one time unit, each R site may undergo an SSE reaction with a
probability $p_{SSE}$. In neutral or acidic medium ($N_{exc}\geq 0$), reaction
(\ref{dissolacidic})  is represented by
\begin{equation}
R\to A ,
\label{reacRA}
\end{equation}
and in basic medium  ($N_{exc}<0$), reaction (\ref{dissolbasic})  is represented
by
\begin{equation}
R+B(nn) \to P+E(nn) 
\label{reacRBPE}
\end{equation}
(here $(nn)$ refers to a B site which is neareast neighbor of
the R site).
Any of the above anodic reactions is possible only if there is another surface
site (R or P) which can mediate the associated cathodic process. This
corresponds to the
electrochemical reactions (\ref{cathodicacidic}-\ref{cathodicbasic}), and is
represented in the model by
\begin{equation}
S + A(nn) \to S + E(nn) ,
\label{reacSASE}
\end{equation}
or
\begin{equation}
S+E(nn) \to S+B(nn) . 
\label{reacSESB}
\end{equation}
If none of the S sites has a nearest neighbor A or E, then the cathodic reaction
is
impossible, and consequently, the anodic one does not occur.

Subsequently, the possibility of local reactions
is examined in a random sweep of the lists of R and P sites.  The
covering of the metal by the passive layer in neutral or basic medium is
represented by
\begin{equation}
R\to P .
\label{reacRP}
\end{equation}
We assume that it occurs with probability $1$ in basic medium ($N_{exc}<0$) and
probability $p_{cor1}$ in neutral medium ($N_{exc}=0$). On the other hand, in
acidic medium ($N_{exc}>0$), R sites may be immediately dissolved, which
corresponds to the reaction
\begin{equation}
R\to E .
\label{reacRE}
\end{equation}
In order to account for the effects of increasing acidity, this process is
assumed to occur with probability $p_{cor2}N_{exc}$, where $p_{cor2}$ is
constant. Here we assume that $p_{cor2}\ll 1$, following the idea that R sites
are preferably dissolved in an anodic reaction.

The last set of possible transformations in a time unit account for
depassivation. It is assumed that this process is not possible in a basic medium 
($N_{exc}<0$) and rather slow in a neutral medium, so that the transformation
\begin{equation}
P\to E 
\label{reacPE}
\end{equation}
occurs with a small probability $p_{oxi}$ when $N_{exc}=0$. On the other hand,
in order to represent the effect of aggressive anions (reaction
\ref{dissolaggressive}) in acidic medium ($N_{exc}>0$), we assume that
(\ref{reacPE}) occurs with probability ${p'}_{oxi}N_{exc}$, where ${p'}_{oxi}$
is constant, but not very small. In previous works on the model
\cite{vautrin1,stafiej}, a fixed value ${p'}_{oxi}=1/4$ was
used. However, this process plays an essential role in the crossover from slow
to rapid corrosion, thus possible variations of its rate will be considered
here.

Finally, to represent diffusion in the solution, the random walk of particles
A and B is considered. During one time unit,
each A and B particle tries to perform $N_{diff}$ steps to nearest neighbor
sites. If the step takes that particle to a site with neutral solution (E), then
the particles exchange their positions:
$A_1+E_2\to E_1+A_2$ or $B_1+E_2\to E_1+B_2$, where indexes $1$ and $2$ refer to
neighboring lattice sites. If the step takes an A particle to a
site with a B particle or vice versa, then they are annihilated and both are
replaced by E particles: $A_1+B_2\to E_1+E_2$. This represents the
reestablishment of neutrality by diffusion of the local excess of $H^+$ and
${OH}^-$ and by mutual irreversible neutralization.
In all other cases A and B particles  remain at their position.

It is important to recall that the probabilities defined above, as well as
$N_{diff}$, can be viewed as the time rates for the corresponding processes. If
the lattice parameter is $a$, then the diffusion coefficient of the A and B
particles is $a^2N_{diff}/2d$, where $d$ is the system dimensionality.
To simplify and reduce the number of parameters we take equal diffusion
coefficients 
for both species.

\subsection{Initial conditions and values of the model parameters}
\label{parameters}

Although this model can be investigated with several different initial
conditions, one of the most interesting cases is the study of the corrosion
process taking place after a small damage of a protective layer.
This is the case when a painted metal surface is locally damaged, which puts the
metal in contact with an aggressive environment. It frequently occurs in
relatively cheap (not stainless steel) painted materials in their natural
environment, thus it is of great practical interest to investigate corrosion in
such conditions.

In the model, such damage can be represented by the passivation of two sites
of the top layer of a metallic matrix, while it is assumed that all sites
around the metal are inert, i. e. they cannot be dissolved. This initial
condition is illustrated in Fig. 1a and will be explored by a scaling theory
and numerical methods in the next sections.

The most rapid process in this corrosion problem is expected to be that for
passivation of reactive regions in a neutral or basic medium (reaction
\ref{reacRP}). For this reason, the unit rate is associated to this event in
the basic medium, and $p_{cor1}\lesssim 1$ is adopted. On the other hand, the
passivation in acidic medium is difficult, which justifies the choice
$p_{cor2}\ll 1$. In the simulations of this paper, we will consider
$p_{cor1}=0.9$, and $p_{cor2}=0.02$ or $p_{cor2}=0.005$.

The slowest process is expected to be the dissolution of passivated regions in
a neutral medium. Since it occurs with probability $p_{oxi}$ per unit time, the
characteristic time of this process is $\tau \equiv 1/p_{oxi}\gg 1$. However,
dissolution is enhanced in acidic medium, which justifies the use of
${p'}_{oxi} \gg p_{oxi}$.

It is also reasonable to assume that $p_{SSE}\ll 1$, since this is the fraction
of reactions which
generate significant pH inhomogeneities. However, this is the mechanism
responsible for the onset of high corrosion rates after an incubation time,
which suggests that $p_{SSE}\ll p_{oxi}$. Otherwise, the effective rate of metal
dissolution would not be affected by the pH inhomogeneities.

Finally, diffusion rates of particles A and B are also of order $1$ or larger,
which accounts for rapid cancellation of pH fluctuations in the solution.

\subsection{Summary of previous results on the model}
\label{previous}

The above model without the SSE reactions was formerly considered in Ref.
\protect\cite{vautrin}, where the evolution of the corrosion front was analyzed
via simulations and a mean-field approach. The first study of the corrosion
model with SSE reactions was presented in Ref. \protect\cite{vautrin1}, where
the formation of domains of A and B particles in a highly irregular dissolved
region was observed, as well as the existence of a nucleation regime before the
rapid pit propagation.

The first study which focused the nucleation regime was presented in Ref.
\protect\cite{stafiej}, where the distribution of incubation times was obtained
from simulation. It was shown that, at short times, the size of
the dissolved region slowly increased because the rates of dissolution of P
particles were very small and the repassivation of reactive sites was very
frequent. The products of anodic and cathodic reactions create inhomogeneities
in the local pH of the solution, represented by A and B particles, but these
particles mutually neutralize in an $ A+B \rightarrow 0$ reaction. These
features characterize the nucleation regime.

However, anodic and cathodic reactions may occur at two distant points, which
makes the neutralization slow, while the regions enriched in anions are noxious
for repassivation, promoting further dissolution of the metal. Consequently,
after a certain time the size of the cavity and the number of A and B particles
rapidly increase. This corresponds to the onset of pitting corrosion. The
distribution of incubation times characterizing the transition from the
nucleation regime to pitting corrosion was obtained in Ref.
\protect\cite{stafiej} for two different values of the model parameters. It was
shown that the average incubation time decreased as $p_{oxi}$ increased, but no
the study of the relations between incubation time and the other model
parameters was presented there.

\section{Scaling theory}
\label{scaling}

Starting from the configuration in Fig. 1a, one of the initial P sites is
dissolved after a characteristic time of order $\tau/2$ (reaction
\ref{reacPE}). It leads to the appearance of reactive sites at its
neighborhood, as illustrated in Fig. 1b for the case of depassivation of the P
site at the left.

When $p_{SSE}=0$, the subsequent steps of the process are the passivation of
the reactive sites, which leads to the configuration shown in Fig. 1c.
Consequently, an additional time of order $\tau/3$ will
be necessary for dissolution of a region of the passive layer in Fig. 1c and the
onset of new reactive sites. After that, the new R sites will also be rapidly
passivated. Thus, for small $p_{oxi}$ (large $\tau$), the corrosion process is
always slow. A mean-field theory is able to predict its long-term behavior
\cite{vautrin}.

The case of nonzero but small $p_{SSE}$ is much more complex due to the
appearance of the pH inhomogeneities in the solution. Fig. 1d illustrates the
result of an SSE reaction after the configuration of Fig. 1b, followed by
diffusion which leads to the annihilation of the pair AB.
Simulation of the model \cite{stafiej} shows that, at short times, the size of
the dissolved region increases and the region acquires the approximate shape of
a semicircle centered at the point of the initial damage. After a certain
time, the rate of production of A and B particles remarkably increases,
compensating the annihilation effects. This is mainly a consequence of
the enhanced dissolution and the impossibility of passivation in acidic
regions. Consequently, the total rate of dissolution of the metal becomes much
larger and the cavity develops an irregular shape. The time evolution of the
area of the dissolved region (number of dissolved sites) obtained in simulation
is illustrated in Fig. 2.

In order to relate the geometrical features of the dissolved cavity with the
model parameters, we will consider that it has the semicircular shape of radius
$R$ shown in Fig. 3. In the following, we will  estimate the characteristic
times for creation of new A and B particles in that cavity and the
characteristic time for the annihilation of one of those pairs. Matching these
time scales, we can predict conditions for the incubation to occur.
This reasoning follows the same lines which are successfully applied to study
the crossover between different growth kinetics in Ref. \protect\cite{rdcor}.

First we consider the mechanisms for depassivation followed by SSE reactions.
There are two possible paths for the dissolution of particles P and creation of
new reactive sites: dissolution in a neutral environment, with probability
$p_{oxi}$, and dissolution in acidic medium, i. e. in contact with A particles,
which typically occurs with probability ${p'}_{oxi}$ (contact with a
single A particle). In any case, the generation of a new particle A requires a 
subsequent SSE reaction.

In the case of dissolution in a neutral environment, the average
time necessary for dissolution of a single P particle is $\tau/N_P$, where $N_P$
is the current number of P particles at the surface of the cavity.
This is illustrated in Fig. 4a. After
dissolution, one or two R particles are generated, and each one may undergo an
SSE reaction with probability $p_{SSE}$. Otherwise, these R particles are
rapidly passivated, and new SSE reactions will be possible only after another
depassivation event. These two possibilities are also illustrated in Fig. 4a.
Thus, the average time for creation of
a pair AB inside the cavity from this path is of order $\tau_{cre}\sim
\tau/N_P/p_{SSE}$. While $\tau$ must be interpreted as a time
interval, here $p_{SSE}$ must be viewed as a dimensionless probability which
indicates the fraction of depassivation events that are followed by SSE
reactions. Now, since $N_P$ is of the order of the number of sites at the
surface of the cavity, we have
\begin{equation}
N_P\sim \pi R/a
\label{np}
\end{equation}
for the cavity of radius $R$ (Fig. 3 - $a$ is the lattice parameter). This leads
to
\begin{equation}
\tau_{cre}\sim \frac{\tau a}{\pi R p_{SSE}} .
\label{taucre}
\end{equation}

In the case of dissolution in an acidic environment, we assume that there is
an A particle present in the cavity. The fraction of the time it spends in the
surface of the cavity is approximately the ratio between the number of perimeter
sites and the total number of sites in the cavity (Fig. 3):
$\left( \pi R/a\right) /\left( \pi R^2/2a^2\right) = 2a/R$. The
probability of dissolution of P in contact with a single A is ${p'}_{oxi}$, thus
the characteristic time for dissolution is  $1/{p'}_{oxi}$ in this case. This
process is illustrated in Fig. 4b. Again, the subsequent processes may be SSE
reactions or passivation of the R particles (see Fig. 4b). Thus, the average
time for production of a new A particle is
\begin{equation}
{\tau '}_{cre}\sim \frac{1}{2a/R} \frac{1}{{p'}_{oxi}} \frac{1}{p_{SSE}} =
\frac{R}{2a{p'}_{oxi} p_{SSE}} 
\label{taucre1}
\end{equation}
Again, in this case $p_{SSE}$ must be viewed as a dimensionless fraction of SSE
reactions.

Certainly the above estimates of $\tau_{cre}$ and ${\tau '}_{cre}$ contain
inexact numerical factors, but the dependence on the model parameters is
expected to be captured. However, it is interesting to stress that the
superestimation of ${\tau '}_{cre}$ not only comes from geometrical aspects but
also from diffusion. Indeed, the random movement of the A particle tends to be
slower in contact with the surface, where the number of empty neighbors is
smaller. This increases the average time of A in contact with P. 

Now we estimate the characteristic time for annihilation of a pair of particles
A and B, which represents the smoothing out of local pH inhomogeneities and
leads 
to a decrease of the dissolution rate. Such a pair is illustrated
in Fig. 3. In a first approximation we expect that the average time for
its annihilation, $\tau_{ann}$, is of the order of the number of sites inside
the cavity divided by the number of steps per time unit, $N_{diff}$. Thus
\begin{equation}
\tau_{ann}\sim \frac{\left( R^2/2a^2\right)}{N_{diff}} .
\label{tauann}
\end{equation} 

Notice that, when the cavity is small, $\tau_{ann}\ll
{\tau '}_{cre}\ll \tau_{cre}$,
thus all pairs AB are rapidly annihilated after their production. It means that
the solution is neutral during most of the time. This conclusion is realistic
because, at large length scales, the solution is expected to always be neutral.
However, as $R$ grows, the number of pairs AB in the cavity increases, which
leads to a more rapid dissolution of the metal and slower smoothing out the of
pH
inhomogeneities. For sufficiently low probabilities $p_{oxi}$ of dissolution in
a
neutral medium, as discussed in Sec. \ref{parameters}, we expect that 
$\tau_{cre}$ is always large compared to the other time scales. Thus,
a series of reactions appears when dissolution in acidic medium becomes
rapid enough to counterbalance the neutralizing effect of diffusion. At
this time, if a new pair AB is created, it will induce the creation
of other pairs and, consequently, a succession of SSE reactions. This is
clear from Fig. 4b, in which we observe the production of two A particles in a
small region of the lattice. Thus, at the incubation time, we expect that
$\tau_{ann}\sim {\tau '}_{cre}$. Using Eqs. (\ref{taucre1}) and (\ref{tauann}),
we obtain an average radius
\begin{equation}
R_I \sim \frac{N_{diff}}{\pi p_{SSE} {p'}_{oxi} } \qquad ,\qquad d=2 ,
\label{ri}
\end{equation}
where the index $I$ refer to incubation time and the spatial dimension $d$ is
emphasized.

This result was derived for the model in two dimensions because our simulations
were performed in these conditions. In a three-dimensional system, the
dependence of ${\tau '}_{cre}$ on $R$ does not change, since it is derived from
a boundary to volume ratio. However, $\tau_{ann}$ may increase as $R^3$ instead
of $R^2$ for a three-dimensional cavity, since that is the order of the number
of sites. Consequently, Eq. (\ref{ri}) is expected to be changed to
\begin{equation}
R_I \sim \sqrt{ \frac{N_{diff}}{p_{SSE} {p'}_{oxi}} } \qquad ,\qquad d=3 .
\label{ri3d}
\end{equation}
Thus, independently of the system dimensionality, we observe that $R_I$
decreases as
$p_{SSE}$ or ${p'}_{oxi}$ increase.

One of the interesting points of these results is their independence on
the lattice structure, which is important for comparisons between such a
mesoscopic model and experimental results. In the following, we show that
simulation data are in a good agreement with the predictions in two dimensions
in the cases where the sizes of the cavities and incubation times are large
enough for such a continuous description to apply. This adds more confidence to
the extension of this theoretical analysis to three-dimensional systems.

\section{Simulation results}
\label{simulation}

Our simulations are performed in a square lattice of vertical and
horizontal sizes $L=1000$, with the central sites of the top layer (largest $y$)
initially
labeled as P and the rest labeled as M (Fig. 1a). The process is interrupted
before or when the corrosion front reaches the sides or bottom of the box.
Typically one hundred independent runs have been performed for each set of the
model parameters.

Following previous experience with simulation of this model, the incubation time
is defined as that in which the number of particles A (or B) is 20. Although
our scaling theory predicted the time necessary for the creation of the second
A particle, it is observed that within a very narrow time interval the number
of A and B particles rapidly increases from values near 2
or 3 to some tenths. Thus,
defining the incubation time in the presence of a smaller numbers of particles
A, such as 10 or 15, leaded to negligible changes in the final estimates of
incubation times and radii of the cavities.

The first test of the relation (\ref{ri}) addresses the dependence of $R_I$ on
$p_{SSE}$. Simulations for $0.01\leq p_{SSE}\leq 0.1$ are performed,
considering fixed values of the other parameters: $p_{oxi}={10}^{-3}$,
${p'}_{oxi}=0.25$, $p_{cor1}=0.9$, $p_{cor2}=0.02$, $N_{diff}=1$. In Fig. 5 we
show a log-log plot of $R_I$ versus  $p_{SSE}$ with a linear fit of slope
$-1.05$. This is in a good agreement with Eq. (\ref{ri}) for
constant ${p'}_{oxi}$ and $N_{diff}$, which suggests $R_I\sim 1/p_{SSE}$. This
is certainly the most important test to be performed with this model because it
is much more difficult to estimate a relative probability of spatially
separated reactions in an experiment than to measure diffusion coefficients or
dissolution rates.

We also test the predicted dependence of $R_I$ on $N_{diff}$ by simulations
with the same parameters above, except that we fix $p_{SSE}=0.1$  and
$N_{diff}$ is varied between $1$ and $10$. In Fig. 6 we show a log-log plot of
$R_I$
versus $N_{diff}$ with a linear fit of the data points with larger $N_{diff}$.
That fit gives $R_I\sim {N_{diff}}^{0.7}$, which is a slightly slower dependence
than the linear one predicted by Eq. (\ref{ri}). However, it illustrates the
rapid increase of $R_I$ with the diffusion coefficient, and from Fig. 6 it is
clear that the effective exponent in the relation between $R_I$ and $N_{diff}$
tends to increase as $N_{diff}$ increases \cite{effectiveexp}. The deviation
from the linear relation may be attributed, among other factors, to the
assumption of free random walk properties for A and B particles in our scaling
picture. 

We also confirm that $R_I$ rapidly increases as ${p'}_{oxi}$ decreases,
particularly when the latter is small, as suggested by Eq. (\ref{ri}). We
performed simulations with $p_{oxi}={10}^{-4}$,
${p}_{SSE}=0.01$, $p_{cor1}=0.9$, $p_{cor2}=0.02$, $N_{diff}=1$, and various
${p'}_{oxi}$ between $0.25$ and $0.025$. In Fig. 7 we show a log-log plot of
$R_I$ versus ${p'}_{oxi}$. For small ${p'}_{oxi}$ the incubation radius $R_I$
rapidly increases with decreasing ${p'}_{oxi}$, although the exact dependence
predicted in Eq. (\ref{ri}) is not observed. However, when ${p'}_{oxi}$ is very
small, the average time of creation of new A particles in acidic media (${\tau
'}_{cre}$) becomes very large and possibly comparable to the time for
creation in neutral
media ($\tau_{cre}$), even if $p_{oxi}$ is small. Consequently, a competition
between these mechanisms may appear and rule out the above scaling picture.

Finally, we also test a possible dependence of $R_I$ on $p_{oxi}$. Simulations
are performed with ${p'}_{oxi}=0.25$, $p_{cor1}=0.9$, $p_{cor2}=0.02$,
$N_{diff}=1$ and $p_{SSE}=0.1$, while $p_{oxi}$ varies between ${10}^{-2}$ and
${10}^{-4}$. The estimates of $R_I$ range between $9.67\pm 1.50$
and $9.87\pm 1.48$, i. e. they fluctuate $2\%$ while $p_{oxi}$ varies two
orders of magnitude. This supports our prediction that $R_I$ does not depend on
this rate. On the other hand, as discussed below, the average incubation time
strongly depends on that quantity.

Another interesting point of our scaling analysis is the possibility of
obtaining reliable estimates of the order of magnitude of the radius of the
dissolved region at the incubation time. For instance, simulations with
$p_{oxi}={10}^{-3}$, ${p'}_{oxi}=0.25$, $p_{cor1}=0.9$, $p_{cor2}=0.02$,
$N_{diff}=1$ and $p_{SSE}=0.01$ gave $R_I/a=53\pm 8$, while Eq. (\ref{ri}) gives
$R_I=127$. 

\section{Incubation time}

Here we  derive relations for the average incubation time by extending the
analysis that leads to the scaling theory of Sec. \ref{scaling}. Again, we 
focus on the ranges of realistic values of the model parameters presented in
Sec. \ref{parameters}.

The initial configuration of the system, shown in Fig. 1a, evolves to that shown
in Fig. 1b after an average time $\tau/2$. The latter has a large probability
[${\left( 1-p_{SSE}\right)}^2$] of being passivated, which
gives rise to the configuration with $3$ particles P of Fig. 1c. Otherwise, with
a small probability of order $2p_{SSE}$, one of the R particles undergoes an SSE
reaction, increasing the size of the dissolved region, as shown in Fig. 1d.
However,
due to the diffusion of A and B and the high probability of passivation of new
R sites, there is a large probability that the first configuration of Fig. 1d
evolves to passivated configurations with two or three E particles and four or
five P particles. For instance, for $p_{sse}=0.1$, we estimate
that nearly $80\%$ of the initial configurations will evolve to the passivated
state of Fig. 1c, nearly $10\%$ will evolve to passivated states with four P
and two E particles (last configuration of Fig. 1d with R replaced by P), and
nearly $10\%$ will evolve to passivated states with five P and three E
particles. The probability that A and B particles of the configuration in Fig.
1d survive and their number eventually increases is negligible.

Notice that the time intervals for diffusion of A and B particles and for the
SSE reactions before passivation are both very small compared to
$\tau$. Thus, the new passivated configurations described above are obtained
after a time which is also approximately $\tau/2$, corresponding to the first
dissolution event (process from Fig. 1a to Fig. 1b).

After repassivation of the surface, one of the P particles will be dissolved 
after a characteristic time of order $\tau/N_P$. From the above discussion, $N_P
=3$ is the most probable value for small $p_{SSE}$ (Fig. 1b), although $N_P=4$
and $N_P=5$ have non-negligible probabilities of occurrence. Thus, new
configurations with reactive sites will appear after the total time
$\tau/2+\tau/N_P\left( 2\right)$, where ${N_P\left( 2\right)}$ is an average
number of P particles before the second dissolution event
(in neutral media). From the above discussion we see that ${N_P\left( 2\right)}$
is slightly
larger than 3 for small $p_{SSE}$ (${N_P\left( 2\right)}=3$ for $p_{SSE}=0$).

Subsequently, passivation of reactive sites, diffusion and SSE reactions 
take place, but in all cases the time intervals are much smaller than
$\tau$. New passive configurations will be generated, with larger $N_P$ and
consequently, characteristic times $\tau/N_P$ to be depassivated (dissolution
of one P). Thus, the average incubation time is expected to have the general
form
\begin{equation}
\langle \tau_{INC} \rangle \approx \frac{\tau}{2} + \frac{\tau}
{N_P\left( 2\right)} +
\frac{\tau}{N_P\left( 3\right)} + \dots + \frac{\tau}{{N_P}^{INC}} .
\label{tauinc}
\end{equation}
Here, $N_P\left( i\right)$ is the average number of P particles in the
passivated
states just before the $i^{th}$ dissolution event, and ${N_P}^{INC}$ is
the number of P particles at the time in which the rapid dissolution begins, i.
e. at the incubation time. From Eq. (\ref{np}), we have ${N_P}^{INC}\sim \pi
R_I/a$, thus $R_I$ will determine the term where the series in Eq.
(\ref{tauinc}) will be truncated.

This series resembles a harmonic series which is truncated at a certain term.
Indeed, for very small $p_{SSE}$, the subsequent terms are expected to be close
to consecutive integer values. Moreover, for $p_{SSE}=0$ the harmonic series is
recovered and $\langle \tau_{INC} \rangle$ diverges, which is consistent with
the fact that there is no incubation process without the SSE reactions
\cite{vautrin}.

One of the interesting features of the expansion in Eq. (\ref{tauinc}) is that
it allows for the separation of the effects of the main chemical reactions and
the
effects of dissolution in neutral environment. While the former are important
to determine $R_I$ (Eq. \ref{ri}) and, consequently, the number of terms in the
expansion, the latter determine the value of the characteristic time $\tau$. On
the other hand, the sensitivity of $\langle \tau_{INC} \rangle$ on variations
of $R_I$ is not high because the main contributions to the sum in Eq.
(\ref{tauinc}) come from the terms with small $N_P$. Thus, significant
variations in the incubation time are only expected from variations in the
probability of dissolution in neutral medium, $p_{oxi}=1/\tau$. This is a
somewhat expected feature because $\tau$ is the largest characteristic time of
the relevant events in this system.

The above analysis is fully supported by our simulation data. In Fig. 8 we show
a log-log plot of $\langle \tau_{INC} \rangle$ versus $\tau\equiv 1/p_{oxi}$,
obtained in simulations with fixed values of the other parameters:
${p'}_{oxi}=0.25$, $p_{cor1}=0.9$, $p_{cor2}=0.02$, $N_{diff}=1$, $p_{SSE}=0.1$.
The linear fit of the data, shown in Fig. 8, has a slope $0.96$, which is very
close to the value $1$ expected from Eq.
(\ref{tauinc}) with constant denominators in all terms.

The above results imply that $\langle \tau_{INC} \rangle$ decreases as
$p_{oxi}$ increases.
Simulations of the model with different values of ${p'}_{oxi}$, keeping the
other parameters fixed, also show a decrease of $\langle \tau_{INC} \rangle$ as
${p'}_{oxi}$ increases. On the other hand, both oxidation probabilities are
expected to be enhanced by increasing the concentration of aggressive anions in
solution. This parallels the experimental observation of a decrease in the
incubation time as the concentration of those anions increases
\cite{hassan,rehim,amin,zaky}, which shows the reliability of our model.

In Fig. 9 we show 
$\langle \tau_{INC} \rangle$ versus $p_{SSE}$ obtained with the same set of
parameters of the data as in Fig. 5. The observed dependence qualitatively 
agrees with Eq.(\ref{tauinc}): as $p_{SSE}$ increases, the radius $R_I$
decreases 
and, consequently, the number of terms in the series of Eq. (\ref{tauinc}) also
decreases. However, due to the particular structure of Eq.
(\ref{tauinc}), it
is difficult to predict a simple relation between $\langle \tau_{INC} \rangle$
and $p_{SSE}$.

From Eq. (\ref{tauinc}) we are also able to provide the correct order of
magnitude 
of the average incubation time. For instance, considering the simulation with
$p_{oxi}={10}^{-3}$,
${p'}_{oxi}=0.25$, $p_{cor1}=0.9$, $p_{cor2}=0.02$, $N_{diff}=1$ and
$p_{SSE}=0.01$, we obtained $R_I/a\approx 53$ and $\langle \tau_{INC} \rangle
\approx 1.2\times {10}^4$. On the other hand, using Eq. (\ref{np}), that value
of $R_I$ leads to ${N_P}^{INC}\approx 166$. Since $p_{SSE}$ is small, we may
approximate the series in Eq. (\ref{tauinc}) by a harmonic series truncated at
this value of ${N_P}^{INC}$, so that the theoretical estimate of incubation time
is $4.7\times {10}^{3}$, i. e. of the same order of magnitude of the simulation
value.

\section{Relations to other corrosion models and reaction-diffusion systems}
\label{relations}

In our model, the crossover from slow corrosion to a rapid corrosion process
takes place when regions with large number of A and B particles are produced in
the solution. Indeed, previous simulation work on the model has already
shown the presence of spatially separated domains of A and B particles
in the solution \cite{vautrin1} after the incubation time, and this separation
slows down the annihilation of A and B pairs.

This phenomenon resembles the segregation effects
observed in reaction-diffusion systems of the type $A+B\to 0$ in confined
media \cite{wilczek,oz,anacker,lindenberg,lin,kopelman,reigada}, such as
systems with tubular geometries and fractal lattices. In those systems, there
is no injection of new particles and the initial distribution of reactants is
random. However, after a certain time, large domains of A and B particles are
found, and the annihilation process is possible only at the frontiers of
those domains. Consequently, the concentration of A and B particles slowly decay
when compared to the reactions of the type $A+A\to 0$
\cite{anacker,lindenberg}. As far as we know, this phenomenon was not
experimentally observed yet, although the depletion zones of a related model
($A+B\to B$ reactions) were already observed in photobleaching of fluorescein
dye by a focused laser beam \cite{trap1,trap2}.

However, the segregation is not expected for the reaction
$A+B\to 0$ in three dimensions, and in two dimensions it is expected to be
marginal \cite{lindenberg}. This is the case in our model. The production of A
and B particles takes place at the surface of the cavity, and this surface may
play the same role of the rigid boundaries of confined media (restricting the
diffusion of A and B particles) during a short times after their creation.
However, the main mechanism contributing to segregation in our model is the
preferential production of new A particles by dissolution in acidic media, i. e.
production of new A particles close to the other A.
Despite the differences in the mechanisms leading to segregation between our
model and the reaction-diffusion systems of the type $A+B\to 0$, it is clear
that in both cases the mixing of A and B domains is slow due to diffusion
restrictions.

It is also important to notice that our model has significant differences from
those of Refs. \protect\cite{meakin1,meakin2}, which also aim at representing
universal features of corrosion processes. They also consider the interplay
between dissolution and passivation, with the former being limited by diffusion
of
an aggressive species in the solution. The concentration of the aggressive
species depends only on the properties of the solution in contact with the
metal. However, in our model the particles responsible for enhancing the
dissolution (A and B, or pH inhomogeneities) are themselves dissolution
products. This feature implies that the transition from the nucleation stage
to pitting corrosion is a consequence of an auto-catalytic production of those
particles. Instead, the models of Refs. \protect\cite{meakin1,meakin2} and
related stochastic models of pitting corrosion
\cite{nagatani1,nagatani2,nagatani3} were suitable to represent features of the
regime of pit propagation.

\section{Discussion and conclusion}

We studied a model for corrosion and passivation of a metallic surface after a
small damage to its protective layer, in which an initial regime of slow
corrosion crosses over at the incubation time to a regime of rapid corrosion. 
The dramatic increase of the corrosion rate is related to the presence of acidic
regions in the solution and the autocatalytic enhancement of pH inhomogeneities
due to spatially separated anodic and cathodic reactions. Our scaling analysis
of the model is based on the matching of the characteristic
times of creation and neutralization of pH inhomogeneities in the solution and
leads to an estimate of the average radius $R_I$ of the dissolved region
at the incubation time. That radius decreases with the rate of spatially
separated reactions and the rate of dissolution in acidic media, and it
increases with the diffusion coefficient of particles A and B in the
solution, which tells us how fast the suppression of pH inhomogeneities takes
place.
The average incubation time is written as the sum of a series of
characteristic times for the slow dissolution in neutral media. It
has a complex dependence on $R_I$ and linearly increases with the rate of
dissolution in neutral media. These results are confirmed by
numerical simulation in two-dimensional lattices, but the extension of
the theory to three dimensions is also discussed. Relations to other
reaction-diffusion systems with segregation and other corrosion models are
discussed. Since the relative values of
the model parameters are expected to provide a realistic description of real
corrosion processes, we believe that this work may be useful for the analysis
of experimental work in this field.


\vfill\eject

\begin{figure}[!h]
\includegraphics[clip,width=0.80\textwidth, 
height=0.50\textheight,angle=0]{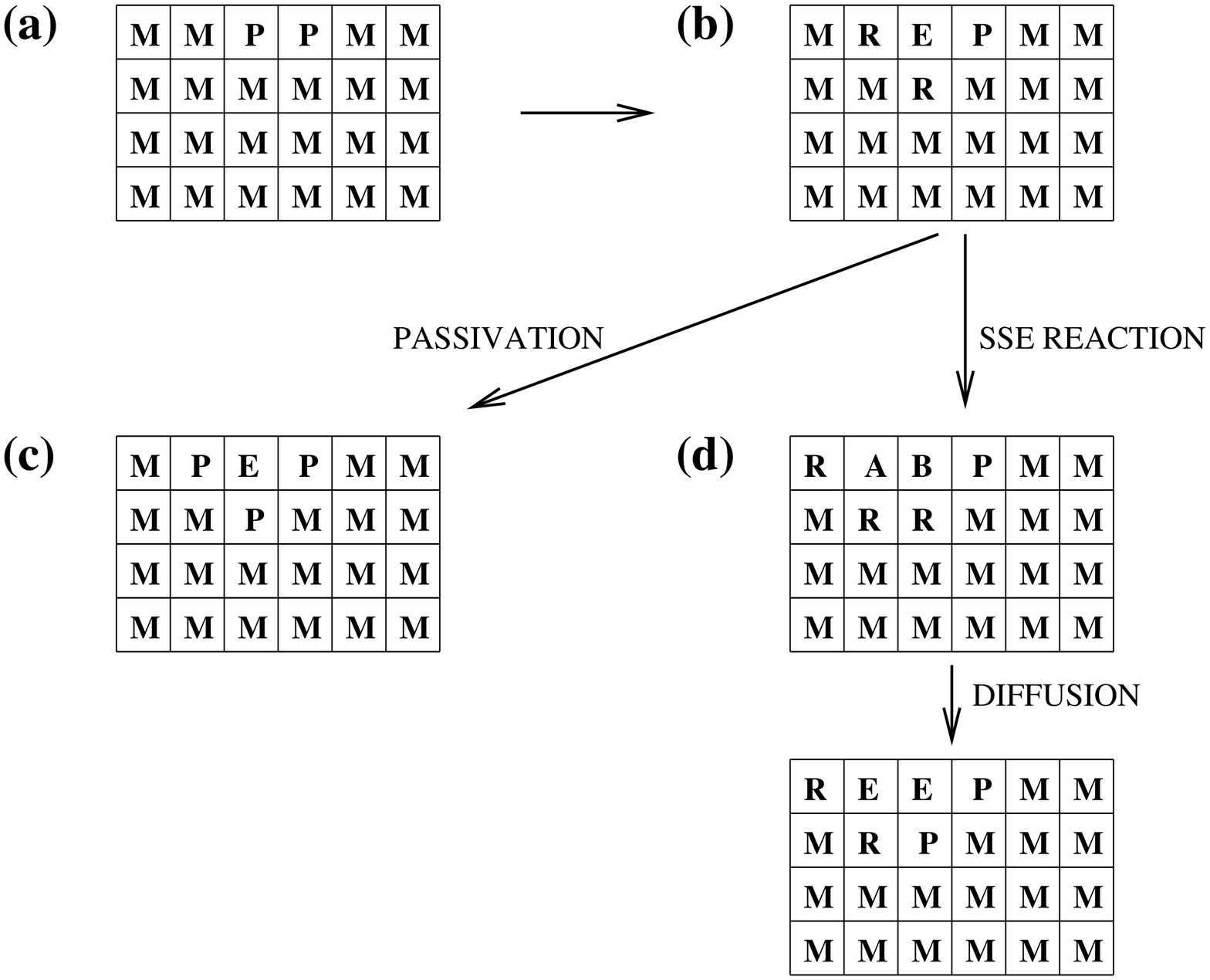}
\caption{\label{fig1} (a) Initial configuration of the lattice, with two
passivated sites in the top layer of the metallic matrix. (b) Configuration
obtained after dissolution of the left P particle. (c) Configuration obtained
after passivation of the R sites. (d) Configurations obtained after the leftmost
R of (b) suffered an SSE reaction and after diffusion and annihilation of A and
B.}
\end{figure}

\begin{figure}[!h]
\includegraphics[clip,width=0.60\textwidth,
height=0.42\textheight,angle=0]{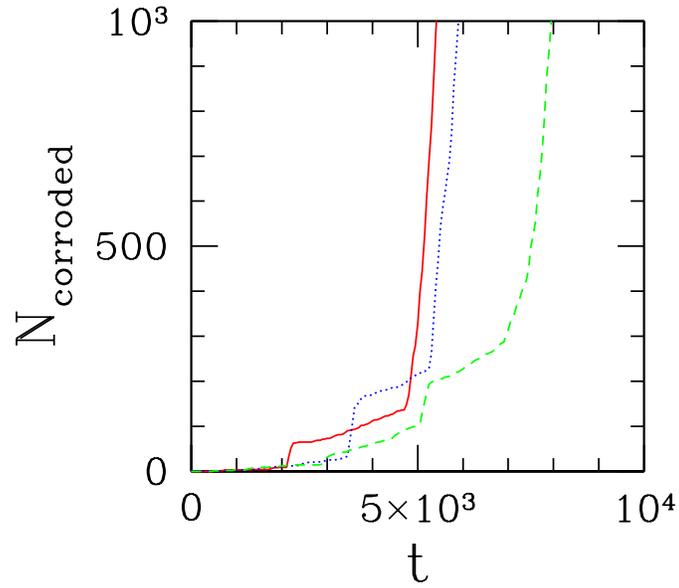}
\caption{\label{fig2} Number of dissolved sites as a function of time, obtained
in three different realizations of the model. Parameter values are
$p_{SSE}=0.02$, $p_{oxi}={10}^{-3}$,
${p'}_{oxi}=0.25$, $p_{cor1}=0.9$, $p_{cor2}=0.02$, and $N_{diff}=1$.}
\end{figure}

\begin{figure}[!h]
\includegraphics[clip,width=0.60\textwidth, 
height=0.225\textheight,angle=0]{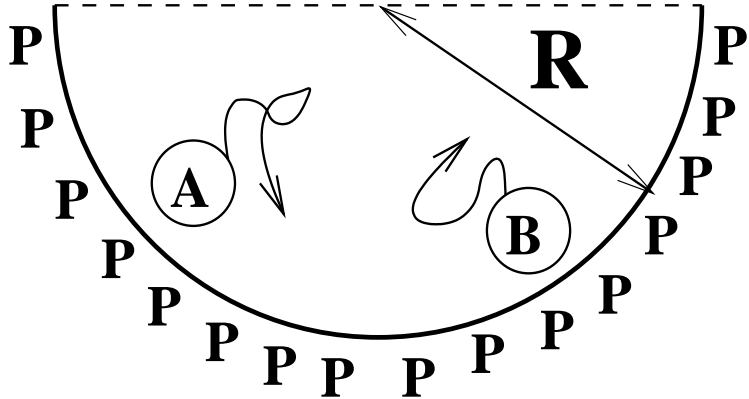}
\caption{\label{fig3} Scheme of the cavity during the incubation process,
assuming a semicircular shape. Most sites of the surface are passivated and
particles A and B are allowed to diffuse in the cavity.}
\end{figure}

\begin{figure}[!h]
\includegraphics[clip,width=0.90\textwidth, 
height=0.7\textheight,angle=0]{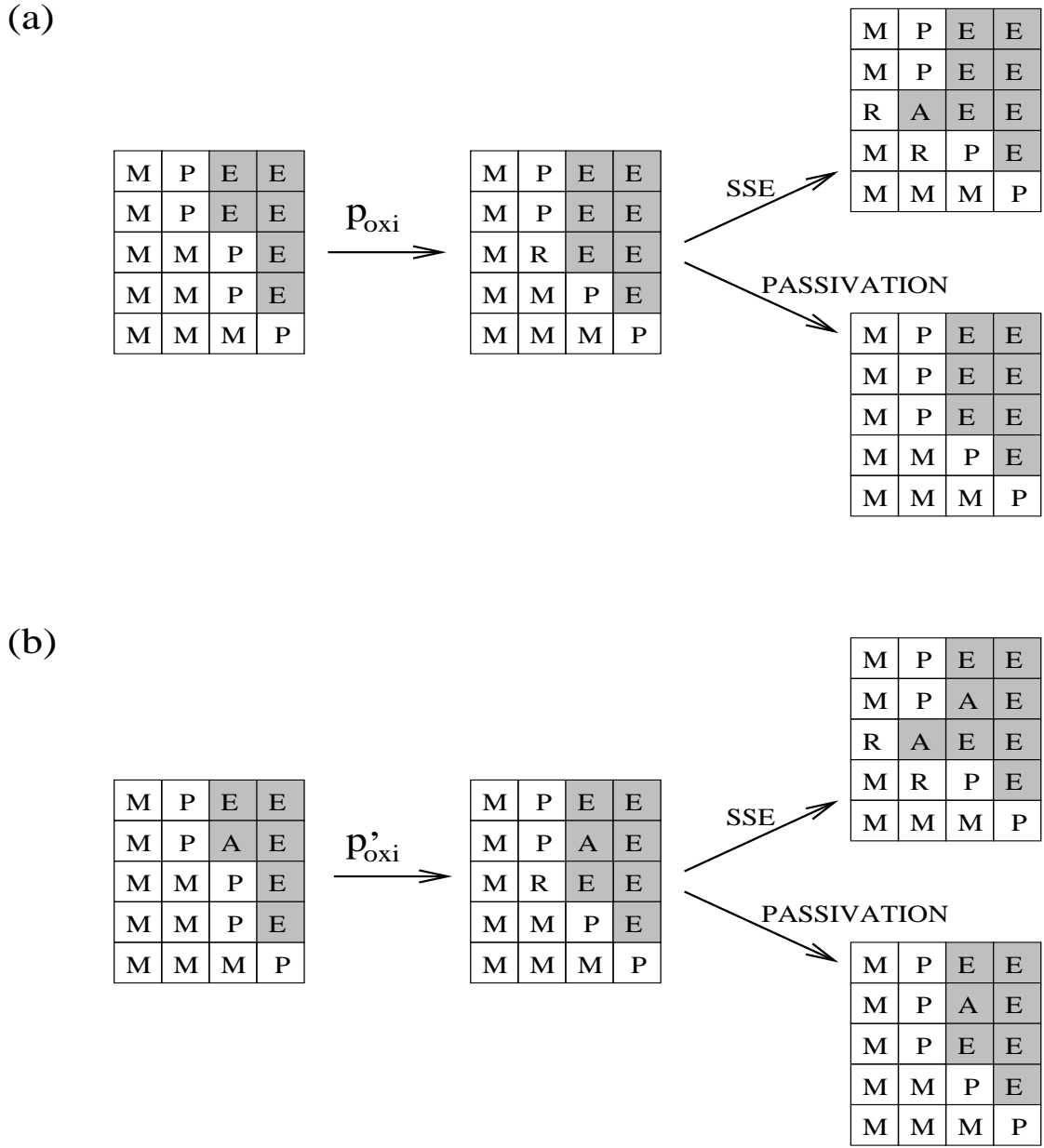}
\caption{\label{fig4} (a) Scheme showing a region of the surface of the cavity
where dissolution in neutral medium occurs, followed by SSE reactions or by
passivation of reactive sites. The solution sites are shaded in order to
illustrate the evolution of the front. (b) The same scheme with dissolution in
acidic medium, i. e. dissolution of a P site with a neighboring A.}
\end{figure}

\begin{figure}[!h]
\includegraphics[clip,width=0.80\textwidth, 
height=0.57\textheight,angle=0]{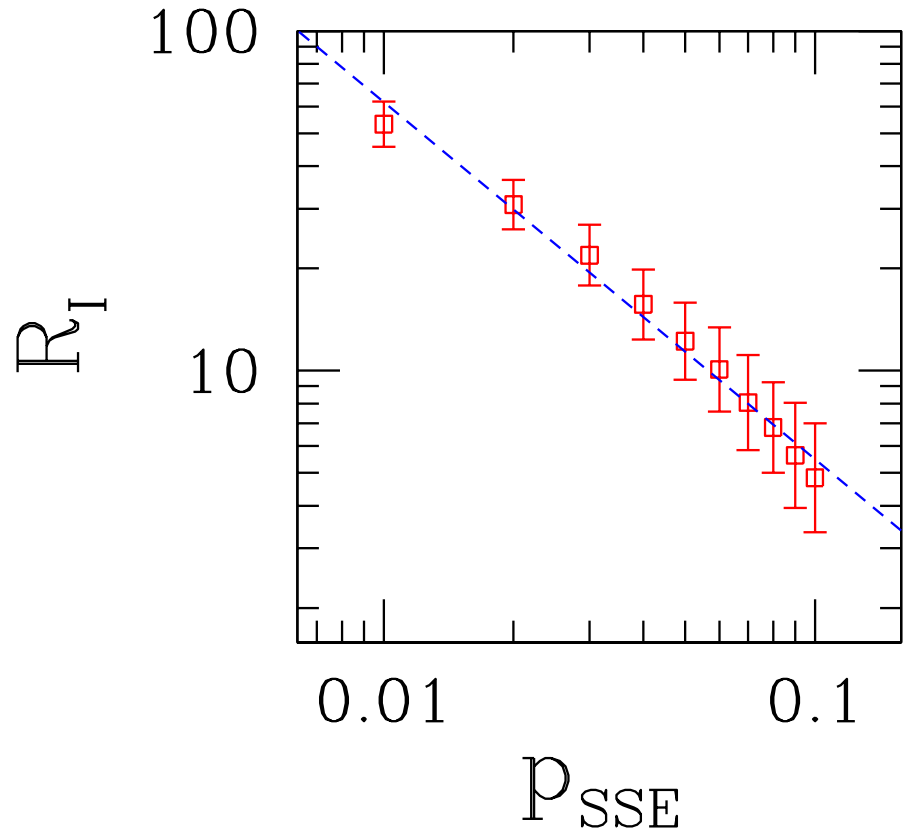}
\caption{\label{fig5} Incubation radius as a function of
$p_{SSE}$, for small $p_{SSE}$, with $p_{oxi}={10}^{-3}$,
${p'}_{oxi}=0.25$, $p_{cor1}=0.9$, $p_{cor2}=0.02$, and $N_{diff}=1$.}
\end{figure}

\begin{figure}[!h]
\includegraphics[clip,width=0.80\textwidth, 
height=0.57\textheight,angle=0]{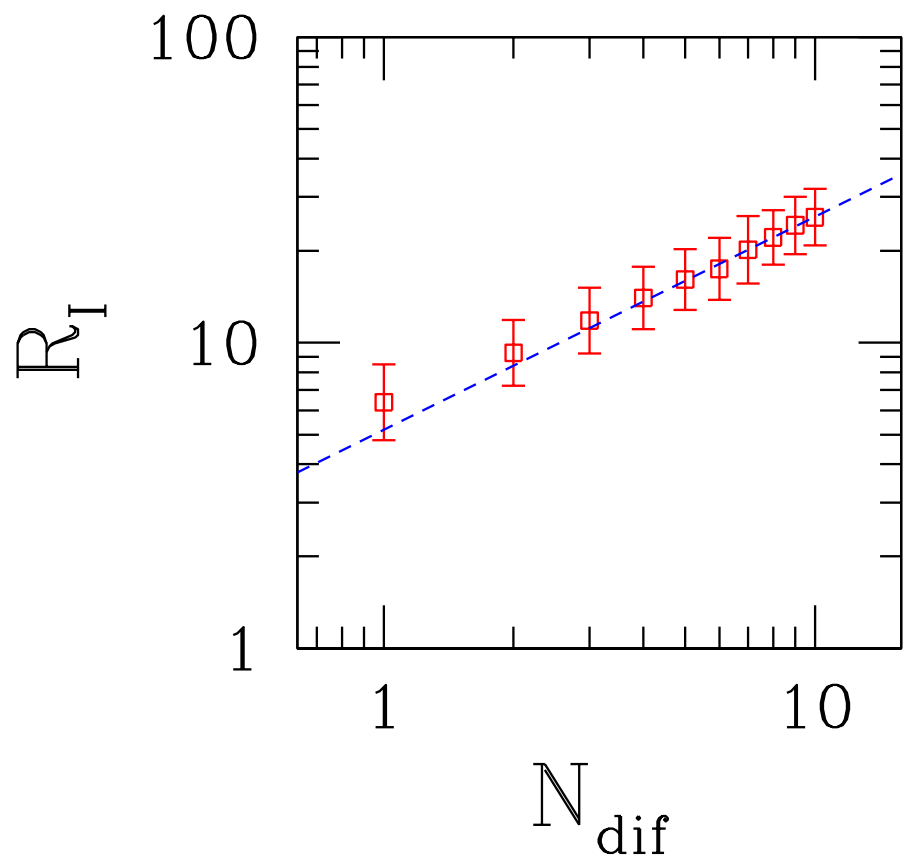}
\caption{\label{fig6} Incubation radius as a function of $N_{diff}$, with
$p_{SSE}=0.1$, $p_{oxi}={10}^{-3}$,
${p'}_{oxi}=0.25$, $p_{cor1}=0.9$, and $p_{cor2}=0.02$.}
\end{figure}

\begin{figure}[!h]
\includegraphics[clip,width=0.80\textwidth, 
height=0.57\textheight,angle=0]{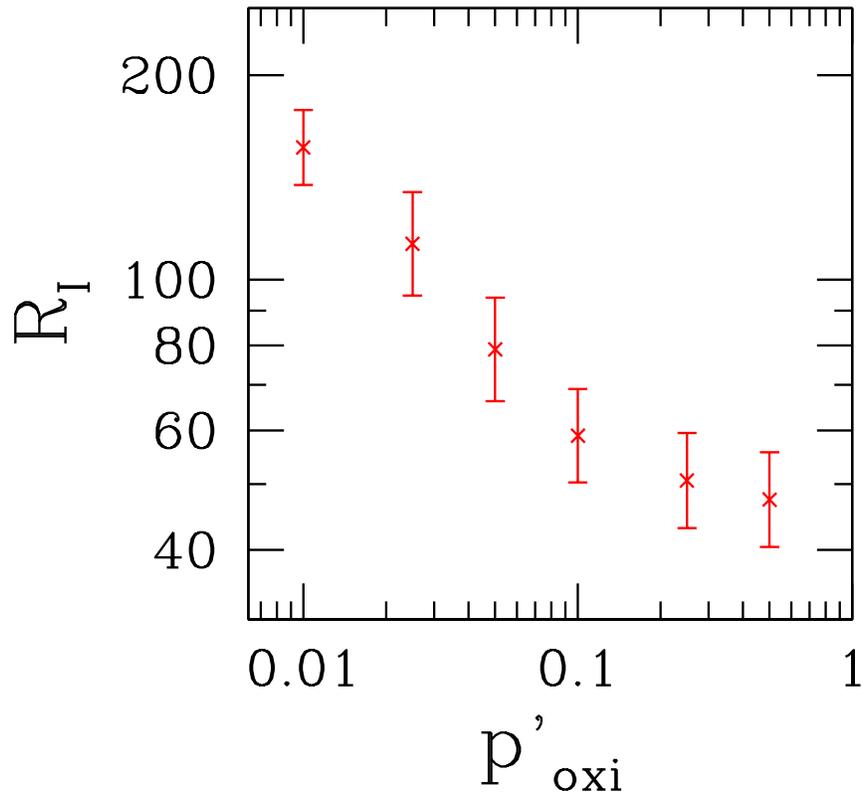}
\caption{\label{fig7} Incubation radius as a function of the probability of
dissolution in acidic media, ${p'}_{oxi}$, with $p_{SSE}=0.1$,
$p_{oxi}={10}^{-3}$, $p_{cor1}=0.9$, $p_{cor2}=0.02$, and $N_{diff}=1$.}
\end{figure}

\begin{figure}[!h]
\includegraphics[clip,width=0.80\textwidth, 
height=0.57\textheight,angle=0]{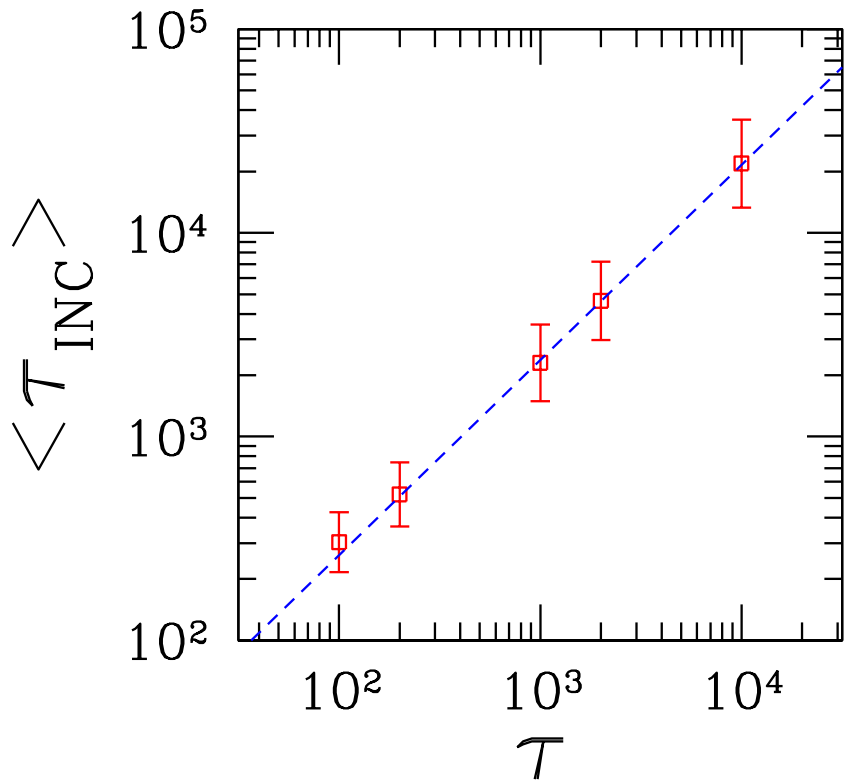}
\caption{\label{fig8} Average incubation time as a function of the oxidation
time $\tau\equiv 1/p_{oxi}$, with $p_{SSE}=0.1$,
${p'}_{oxi}=0.25$, $p_{cor1}=0.9$, $p_{cor2}=0.02$ and $N_{diff}=1$.}
\end{figure}

\begin{figure}[!h]
\includegraphics[clip,width=0.80\textwidth, 
height=0.57\textheight,angle=0]{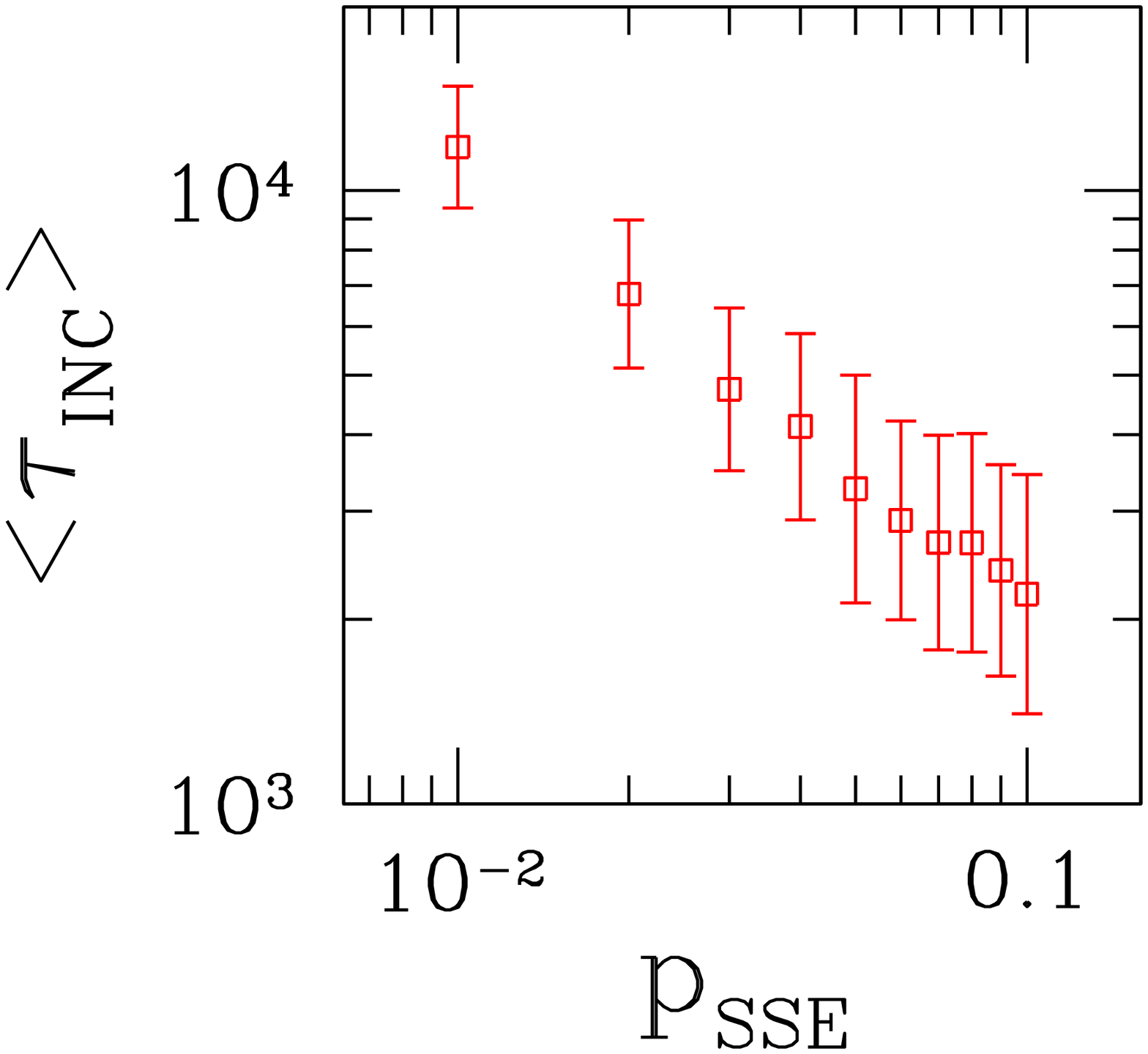}
\caption{\label{fig9}  Average incubation time as a function of $p_{SSE}$,
$p_{oxi}={10}^{-3}$,
${p'}_{oxi}=0.25$, $p_{cor1}=0.9$, $p_{cor2}=0.02$ and $N_{diff}=1$.}
\end{figure}


\begin{references}
\bibitem{effectiveexp}
The effective exponents for a power-law relation between $R_I$ and $N_{diff}$
were estimated along the same lines of Ref. \protect\cite{rdcor}. They showed a
trend to values close but larger than $1$ as $N_{diff}\to\infty$ (the value $1$
corresponds to the linear relation  predicted by Eq. \ref{ri}).
\end{references}
\end{document}